\documentstyle[preprint,aps]{revtex}
\newcommand{\be}{\begin{equation}} \newcommand{\ee}{\end{equation}} 
\newcommand{\bea}{\begin{eqnarray}}\newcommand{\eea}{\end{eqnarray}}

\begin{document}
\draft
\preprint{MRI-PHY/96-20, cond-mat/9607009}
\title{Novel Correlations in Arbitrary Dimensions}
\author{Pijush K. Ghosh$^{*}$}
\address{The Mehta Research Institute of
Mathematics \& Mathematical Physics,\\
Allahabad-211002, INDIA.}
\footnotetext {$\mbox{}^*$ E-mail: 
pijush@mri.ernet.in }  
\maketitle
\begin{abstract} 
We present a new three dimensional many-body Hamiltonian with three-body
and five-body interactions. We obtain the exact ground state as well as
some excited states of this Hamiltonian for arbitrary number of particles.
These exact wave-functions describe a novel correlations. Finally, we 
generalize these three dimensional results to arbitrary higher
dimensions.
\end{abstract}
\narrowtext

\newpage

Recently, there has been a renewed interest in the study of
many-body quantum mechanical systems, like the 
Calogero-Sutherland Model (CSM) and its variants in one dimension
\cite{cs,cs1,pr,hs,ap,khare}. This is primarily because these models
are relevant to
many diverse branches in physics \cite{chart}.
Though, such one dimensional many-body systems
have been studied extensively in the recent literature, nothing much is known
about the appropriate generalization of these models to higher dimensions.
As a promising step towards this, it was pointed out recently \cite{mbs} 
that in two dimensions
there exists novel correlations other than the one used in constructing the
Laughlin's trial wave function \cite{laugh}. The exact ground state
as well as some excited states were also obtained for a model many-body
Hamiltonian,
where this novel correlations can be realized.

The purpose of this letter
is to show that this type of
correlations can be appropriately
generalized
to arbitrary higher dimensions. In particular, we construct a new three
dimensional many-body Hamiltonian with three-body and five-body interactions.
We obtain the exact ground state as well as some excited states of this
Hamiltonian. 
The exact wave-functions of this model describe a novel correlations, which
are an appropriate generalization of the two dimensional correlations
introduced in \cite{mbs}. These correlations can be realized by spinless
bosons as well as fermions.
Finally, all these three dimensional results are generalized to arbitrary
higher dimensions.

The CSM is described by $N$ identical particles confined in a one-body harmonic
oscillator potential or on the rim of a circle, and interacting through
each other via a two-body inverse square potential. The wave function
of this model contains a Jastrow-type factor of the form
$J_{ij}=(x_i -x_j)^\lambda
{\mid x_i - x_j \mid}^\alpha$, where $x_i$ and $x_j$ denote the
particle positions
of the $i$th and the $j$th particle, respectively. The parameters $\alpha$ and
$\lambda$ are related to the
strength of the inverse square two-body potential. The wave functions of CSM
are highly correlated and the nature of correlations  
is encoded in the Jastrow-type factor. The Jastrow-type factor
$J_{ij}$ has two interesting properties :\\
(a) $J_{ij}$ vanishes when the position vectors
of two particles coincide. This ensures that no two particles can occupy
the same position at the same time. Also, the two-body interaction in the
Hamiltonian
has singularities, precisely at these points, i.e., at $x_i = x_j$.\\
(b) $J_{ij}$ picks up a factor $(-1)^\lambda$
under the exchange
of particle indices. Consequently, spinless bosons as well as
fermions can be described by putting $\lambda$ equal to zero
or one, respectively.\\
These two properties of the Jastrow-type factor are the
basic criteria for constructing higher dimensional analogue of the CSM. 
The Laughlin's trial wave-function
for a two
dimensional Hamiltonian, describing spin polarized electrons in the lowest
Landau level with a short-range repulsive interaction, indeed inherits these two
properties \cite{laugh}. In particular, the two dimensional Jastrow-type
factor appearing in Laughlin's trial wave function is, 
$J_{ij}=(z_i - z_j)^\lambda {\mid z_i - z_j \mid}^\alpha$,
where $z_i$ and $z_j$ are the particle positions in the complex coordinates.
However, this is not unique in describing two dimensional many-body
systems. One can also consider a Jastrow-type factor of the form 
$J_{ij}= (z_i^m \bar{z}_j^m - z_j^m \bar{z}_i^m)^\lambda
{\mid z_i^m \bar{z}_j^m - z_j^m \bar{z}_i^m \mid}^\alpha$, where $m$ is an
integer and a `bar' denotes the complex conjugation. $J_{ij}$ vanishes,
whenever the relative angle between the position vectors of any two
particles is zero or a multiple of $\frac{\pi}{m}$. The case $m=1$ was
considered in Ref. \cite{mbs}. Note that for this particular value of $m$,
$z_i \bar{z}_j - z_j \bar{z}_i$ is the magnitude of the cross-product of the
positions vectors $\vec{r}_i$ and $\vec{r}_j$ along the $z$-direction, up
to an overall multiplication factor. Thus, the wave function vanishes,
whenever two
particles are on a line passing through the origin. The interactions in the
Hamiltonian, where this correlation can be realized, have singularities along
the lines $\vec{r}_i \times \vec{r}_j = 0$. The zeroes of $J_{ij}$ are the
singularities of the corresponding Hamiltonian where these correlations can
be realized for arbitrary $m$ \cite{pijush}.

There are model many-body Hamiltonian in three dimensions with two-body
and three-body interactions for which the ground state as well as some
excited states can be written down explicitly \cite{cs2}. Particles described by
such Hamiltonians are either distinguishable or bosons\footnote{The fermionic
wave-functions can be constructed only for $N=3$ and $N=4$ 
particles.}. This is because the Jastrow-type factor used in such cases
is $ J_{ij}=(\vec{r}_i - \vec{r}_j )^2$, and is not antisymmetric
under the exchange of particle coordinates. One can also construct a three
dimensional Jastrow-type factor using the quaternion generalization of the
usual complex coordinates \cite{avb}. This Jastrow-type factor satisfies
the basic criteria
(a) and (b). However, this is not physically interesting, since the quaternion
coordinates are anti-commutating in nature and the wave function becomes
$SU(2)$ valued \cite{avb}.

It is tempting to look
for a three dimensional generalization of the new Jastrow-type factor 
introduced in Ref. \cite{mbs}. Naively, one would like to consider,
\be
J_{ij}^b = {\mid \vec{r}_i \times \vec{r}_j \mid}^\alpha, \ \
\vec{J}_{ij}^f =(\vec{r}_i \times \vec{r}_j) {\mid \vec{r}_i
\times \vec{r}_j \mid}^\alpha, 
\ee
\noindent which satisfies the basic criteria. The superscripts `b' and `f'
refer to the bosonic and the fermionic nature of the Jastrow-type
factor, respectively. It is possible to construct many-body Hamiltonian
describing spinless bosons where $J_{ij}^b$ can be realized for arbitrary
number of particles. However, unlike in two dimensions,
$\vec{J}_{ij}^f$ is now a vector. The only way one can construct
fermionic
wave-functions is to use linear combinations of these $\vec{J}_{ij}^f$.
Naturally, the wave-functions become three dimensional vectors. This
is acceptable provided each component of the wave function 
satisfies the same Shr$\ddot{o}$dinger equation
independently, with the same eigen-value \cite{cs2}. In other words, the
fermionic wave-function should span the space of degenerate
states. Unfortunately, this type of correlations can be realized
for only three particles \cite{cs2}.

Thus, in order to find an analogue of the two dimensional
novel correlations in three dimensions, we have to
construct a (pseudo-)scalar using the minimum number of position vectors
such that
it obeys the basic properties (a) and (b) of any Jastrow-type factor. 
However, the lesson from the two dimensional example is to modify
the property (a). In particular, the three dimensional Jastrow-type factor can
vanish, not only at points $\vec{r}_i - \vec{r}_j=0$ or 
lines $\vec{r}_i \times \vec{r}_j =0$, but also on
planes $\vec{r}_i . \vec{r}_j \times \vec{r}_k = 0$. 
Let us define
a three dimensional vector $\vec{Q}_{jk}$ in terms of the position
vectors as,
\be
\vec{Q}_{jk} = \vec{r}_j \times \vec{r}_k .
\label{eq0}
\ee
\noindent The three dimensional
Jastrow-type factor can be constructed by projecting $\vec{Q}_{jk}$ along
the position vector of the $i$th particle. In particular,
\be
{J}_{ijk} = (P_{ijk})^\lambda {\mid P_{ijk} \mid}^\alpha, \ \ 
P_{ijk} = \vec{r}_i . \vec{Q}_{jk} .
\label{eq1}
\ee
\noindent Note that both $\vec{Q}_{jk}$ and $P_{ijk}$ are antisymmetric
under the exchange of particle coordinates. Also, $P_{ijk}$ vanishes
when (i) the relative angle between the $i$-th and the $j$-th particle
is zero or $\pi$ and (ii) the position vectors of any three particles lie
on a plane. The constraint (i) ensures that no two particles can occupy
the same position at the same time. We expect at
this point that if any model many-body Hamiltonian realizes this type of
correlations,
then these will be the conditions for having singularities in the many-body
interactions. Note that the two-dimensional correlations introduced in
\cite{mbs} is a sub-class of (\ref{eq1}).

We now present a many-body system where the correlations (\ref{eq1}) can be
realized explicitly.
Consider the
Hamiltonian,
\be
H = - \frac{1}{2} {\sum_{i=1}^N} \bigtriangledown_i^2 +
\frac{1}{2} {\sum_{i=1}^N} \vec{r}_i^2
+ \frac{g_1}{2}
{\sum_{ R}} \frac{\vec{Q}_{jk}^2}{P_{ijk}^2} + \frac{g_2}{2}
{\sum_{R}} \frac{\vec{Q}_{jk} . \vec{Q}_{lm}}{P_{ijk} P_{ilm}},
\label{eq2}
\ee
\noindent where $R$ denotes the sum over all the indices from $1$ to $N$,
with the restriction
that any two indices can not have the same value simultaneously.
We are working in the units $\bar{h}=m=w=1$, where $h=2 \pi \bar{h}$ is the
Planck's constant, $m$ is the mass of each particle and $w$ is the oscillator
frequency. $g_{1}$ and $g_{2}$ are two dimensionless coupling constants.
In general, these two coupling constants are independent of each other.
However, they get related to each other for the particular set of solutions
we obtain below. The Hamiltonian
is symmetric under the exchange of particle indices. As expected, both the
three-body as well as five-body interactions are singular, in case,
(i) any
two particles are on a line passing through the origin
or (ii) any three particles and the origin of the coordinate system
lie on a plane.

We construct a trial 
wave-function for this Hamiltonian as,
\be
\psi_0 = \prod_{\cal{R}} P_{ijk}^\lambda \ {\mid P_{ijk} \mid}^\alpha
\ exp (- \frac{1}{2} \sum_{i=1}^N \vec{r}_i^2) ,
\label{eq3}
\ee
\noindent where ${\cal{R}} \equiv (i < j < k)$. Note that the
wave-function $\psi_0$ is fermionic in nature for odd $N \geq 3$
only. $\psi_0$ can be considered as bosonic for any $N \geq 3$.
Eq. (\ref{eq3}) is an exact eigen state of (\ref{eq2}) provided
$g_1 = \frac{g}{2} ( \frac{g}{2} -1 )$ and $ g_2 = \frac{g^2}{4}$, where
$g= \alpha + \lambda$. One can solve for $g$ as, 
\be
g = ( 1 \pm \sqrt{1 + 4 g_1}).
\label{eq4}
\ee
\noindent Note that for physical states $g_1 \geq -\frac{1}{4}$.
The solutions in the lower branch are regular only in the limited ranges
$ - \frac{1}{4} \leq g_1 \leq 0$, while in the upper branch the solutions
are regular for $g_1 \geq -\frac{1}{4}$.

The energy corresponding to
$\psi_0$ is given by,
$E_0= \frac{3 N}{2} + \frac{g}{2} N (N-1) (N-2)$.
Note that for $g=0$,
i.e. $g_1=g_2=0$, $E_0$ is exactly the ground state energy for $N$ particles
confined in a three dimensional harmonic oscillator potential.
This is the case for $N=1$ and $2$ also, as the many-body interactions do
not play any role unless $N \geq3$.
The wave-function $\psi_0$ has no nodes other than those corresponding to
the singularities of the three-body and five-body interactions of the
Hamiltonian. Thus, $\psi_0$ is well suited for the ground-state of (\ref{eq2}).
In fact,
it can be shown that $\psi_0$ indeed is the
ground state wave-function by constructing the following anhilation
operators\footnote{One should not confuse $z_i$ in (\ref{eq5}) as the complex
coordinates. $\vec{r}_i$ is defined as $\vec{r}_i= x_i \hat{i} + y_i \hat{j}
+ z_i \hat{k}$, where the unit vectors $\hat{i}$, $\hat{j}$ and $\hat{k}$
span the three dimensional space.},
\bea
& & A_{x_i} = p_{x_i} - i x_i + \frac{ig}{2} \sum_S \frac{(\vec{Q}_{jk})_{x_i}}{
P_{ijk}},\nonumber \\
& & A_{y_i} = p_{y_i} - i y_i - \frac{ig}{2} \sum_S \frac{(\vec{Q}_{jk})_{y_i}}{
P_{ijk}},\nonumber \\
& & A_{z_i} = p_{z_i} - i z_i + \frac{ig}{2} \sum_S \frac{(\vec{Q}_{jk})_{z_i}}{
P_{ijk}},
\label{eq5}
\eea
\noindent where $S$ denotes sum over all the repeated indices from $1$ to $N$,
with the
constraint that any two indices can not have the same value simultaneously.
The Hamiltonian can be written down in terms of these anhilation
operators and the corresponding creation operators as,
\be
H = \frac{1}{2} \sum_{i=1}^N \left [ A_{x_i}^\dagger A_{x_i} +
A_{y_i}^\dagger A_{y_i} +
A_{z_i}^\dagger A_{z_i} \right ] + E_0 .
\label{eq6}
\ee
\noindent The operators $A$'s anhilate the wave function $\psi_0$, and thus
$\psi_0$ is the ground state.

The excited states can be obtained by decomposing the wave-function
$\psi$ as,
\be
\psi(x_i, y_i, z_i) = \psi_0(x_i, y_i,z_i) \phi(x_i, y_i, z_i).
\label{eq7}
\ee
\noindent Plugging the expression (\ref{eq7}) into the Schr$\ddot{o}$dinger
equation
, we have,
\be
\left [ -\frac{1}{2} \sum_{i=1}^N \bigtriangledown_i^2 + \sum_{i=1}^N \vec{r}_i .
\vec{\bigtriangledown}_i - \frac{g}{2} \sum_R \frac{\vec{Q}_{jk} . 
\vec{\bigtriangledown}_i}{P_{ijk}} \right ] \phi = (E_n -E_0) \phi .
\label{eq8}
\ee
\noindent Now if $\phi$ is a function of $t={\sum\atop{i=1}}^N r_i^2$ only,
Eq. (\ref{eq8}) reduces to the confluent hypergeometric equation,
\be
t \frac{d^2 \phi(t)}{dt^2} + [b-t] \frac{d \phi(t)}{dt} - a \phi(t) =0,
\label{eq9}
\ee
\noindent where $b=E_0$ and $a=-\frac{1}{2} ( E_n - E_0 )$. The admissible
solutions of (\ref{eq9}) are the regular confluent hypergeometric functions,
$\phi(t)=M(a,b,t)$. The constant $a$ is determined as $a=-n$ in order to have
normalizable eigen functions, where $n$ is
an integer. Thus, the spectrum is given by $E_n=E_0 + 2 n$. Note that this
spectrum is identical to the CSM as well as to the model Hamiltonian
considered in \cite{mbs}. Unfortunately, $E_n$ is not the complete spectrum
of the Hamiltonian (\ref{eq2}). At present, we do not know how to solve the
Schr$\ddot{o}$dinger equation corresponding to (\ref{eq2}) exactly.

We now show that all these three dimensional results can be generalized
to arbitrary higher dimensions. In $D ( > 2 )$ dimensions, one can construct
a Jastrow-type factor with the help of $D$ position vectors. As a result,
the Hamiltonian contains $D$-body and $(2 D -1)$-body interactions only.
Consider the following $D$-dimensional 
Jastrow-type factor $J_{i_1 i_2 \dots i_D}$,
\bea
& & \vec{Q}_{i_2 i_3 \dots i_D} = \vec{r}_{i_2} \times \vec{r}_{i_3} \times
\dots
\times \vec{r}_{i_D}, \ \
P_{i_1 i_2 \dots i_D} = \vec{r}_{i_1} . \vec{Q}_{i_2 i_3 \dots i_D}
,\nonumber \\
& & J_{i_1 i_2 \dots i_D} = ( P_{i_1 i_2 \dots i_D} )^\lambda {\mid
P_{i_1 i_2 \dots i_D} \mid}^\alpha .
\label{eq10}
\eea
\noindent This $D$ dimensional Jastrow-type factor vanishes whenever
any $p$ particles lie on a $p-1$ dimensional (hyper-)plane\footnote{We
denote one dimensional plane as a line.} passing
through the origin, where
$2 \leq p \leq D$. Note that the novel correlations (\ref{eq10})
can be realized only for $N \geq D$.

Consider the $D$-dimensional many-body Hamiltonian,
\be
H = - \frac{1}{2} \sum_{i=1}^N \bigtriangledown_i^2 + \frac{1}{2}
\sum_{i=1}^N \vec{r}_i^2 + \frac{g_1}{2} {\sum_{R}} \frac{
\vec{Q}_{i_2 i_3 \dots i_D}^2}{P_{i_1 i_2 \dots i_D}^2} + \frac{g_2}{2}
{\sum_{R}} \frac{\vec{Q}_{i_2 i_3 \dots i_D} \times
\vec{Q}_{j_2 j_3 \dots j_D}}{P_{i_1 i_2 \dots i_D} P_{i_1 j_2 \dots j_D}}.
\label{eq10.1}
\ee
\noindent We follow the same summation convention as in (\ref{eq2}).
The eigen-states $\psi_n$ of the Hamiltonian (\ref{eq10.1}) are given
by,
\be
\psi_n = \prod_{\cal R} J_{i_1 i_2 \dots i_D} M(-n, E_0, t) exp (-
\frac{1}{2} \sum_{i=1}^N \vec{r}_i^2),
\label{eq10.2}
\ee
\noindent where the ground state energy 
$E_0 = D \left [ \frac{N}{2} + g \ ^N C_D \right ], \ ^N C_D = \frac{N!}{
D! (N-D)!}$ and ${\cal R} \equiv (i_1 <
i_2 < \dots < i_D)$. The wave-functions $\psi_n$ are bosonic in nature
for arbitrary $N \geq D$. The fermionic description is possible, only
when $^{N-2} C_{D-2}$ is odd. The energy spectrum corresponding to the
eigen-states
(\ref{eq10.2}) is $E_n=E_0 + 2 n$. Note that the ground state energy depends
on the dimensionality of space. However, the difference in energy 
between the $n$-th
excited state and the ground state, i.e. $E_n - E_0$, is independent of $D$.

Eq. (\ref{eq10.2}) is an exact eigen-state of (\ref{eq10.1}) with energy
eigen values $E_n$, provided the dimensionless coupling constants
$g_1$ and $g_2$ are
related to $g$ as,
\bea
& & g_2 = \left ( \frac{g}{(D-1)!} \right )^2, \ \
g_1 = \frac{g}{(D-1)!} \left ( \frac{g}{(D-1)!} - 1 \right ),\nonumber  \\
& & g = \frac{(D-1)!}{2} \left [ 1 \pm \left ( 1 + 4 g_1 \right )
^{\frac{1}{2}} \right ].
\label{eq11}
\eea
\noindent Note that the relation between $g_1$ and $g_2$ is independent
of $D$. Also, the ranges of $g_1$ is
independent of the dimensionality of the space.

In conclusions, we have constructed a $D(>2)$-dimensional many-body
Hamiltonian with
$D$-body and $(2 D - 1)$-body interactions, which realizes a novel
correlations among the spinless bosons as well as fermions. 
We found the ground state and some of the excited states of this model.
The fermionic description of the wave-functions is possible for any
$N \geq D$ obeying the constraint that $^{N-2} C_{D-2}$ is odd. The bosonic
nature of the wave-functions are independent of the total number of
particles. It would be nice if either this model Hamiltonian or the
correlation could be realized in some condensed matter systems.

\acknowledgments

I wish to thank Avinash Khare and M. Sivakumar for many valuable discussions
and, in particular, for drawing my attention to the references \cite{cs2}
and \cite{avb},
respectively.


\begin{references}

\bibitem{cs} F. Calogero, J. Math. Phys. (N.Y.) {\bf 10}, 2191 (1969);
{\bf 10}, 2197 (1969).

\bibitem{cs1} B. Sutherland, J. Math. Phys.(N.Y.) {\bf 12}, 246 (1971);
{\bf 12}, 251 (1971); Phys. Rev. A {\bf 4}, 2019 (1971).

\bibitem{pr} M. A. Olshanetsky and A. M. Perelomov, Phys. Rep. {\bf 71},
314 (1981); {\bf 94}, 6 (1983).

\bibitem{hs} F. D. M. Haldane, Phys. Rev. Lett. {\bf 60}, 635 (1988);
B. S. Shastry, {\it ibid.} {\bf 60}, 639 (1988).

\bibitem{ap} A. P. Polychronakos, Phys. Rev. Lett. {\bf 69}, 703 (1992).

\bibitem{khare} A. Khare, J. Phys. A {\bf 29}, L45 (1996);
D. P. Jatkar and A. Khare, Mod. Phys. Lett A {\bf 11}, 1357 (1996).

\bibitem{chart} B. D. Simons, P. A. Lee and B. L. Altshuler, Phys. Rev.
Lett. {\bf 72}, 64 (1994).

\bibitem{mbs} M. V. N. Murthy, R. K. Bhaduri and D. Sen, Phys. Rev. Lett.
{\bf 76}, 4103 (1996).

\bibitem{laugh} R. B. Laughlin, Phys. Rev. Lett. {\bf 50}, 1395 (1983).

\bibitem{pijush} P. K. Ghosh, unpublished.

\bibitem{cs2} F. Calogero and C. Marchioro, J. Math. Phys. (N.Y.) {\bf 14},
182 (1973).

\bibitem{avb} A. V. Balatsky, Quaternion Generalization of the Laughlin
state and the Three dimensional fractional QHE, cond-mat/9205006.
\end{references}
\end{document}